\documentclass[universe,article,accept,moreauthors,pdftex]{Definitions/mdpi} 

\firstpage{1} 
\pubvolume{xx}
\issuenum{1}
\articlenumber{5}
\pubyear{2020}
\copyrightyear{2020}
\history{Received: 17 August 2020; Accepted: 30 September 2020; Published: date}

\usepackage{natbib}
\usepackage{subcaption}

\Title{Using Unreal Engine to Visualize a Cosmological Volume}

\Author{Christopher Marsden *\orcidA{} and Francesco Shankar}

\AuthorNames{Christopher Marsden, Francesco Shankar}

\address{%
Department of Physics and Astronomy, University of Southampton, Highfield, SO17 1BJ, UK; F.Shankar@soton.ac.uk}

\corres{Correspondence: c.marsden@soton.ac.uk }

\abstract{In this work we present ``Astera'’, a cosmological visualization tool that renders a mock universe in real time using Unreal Engine 4. The large scale structure of the cosmic web is hard to visualize in two dimensions, and a 3D real time projection of this distribution allows for an unprecedented view of the large scale universe, with visually
accurate galaxies placed in a dynamic 3D world. The underlying data
are based on empirical relations assigned using results from N-Body
dark matter simulations, and are matched to galaxies with similar
morphologies and sizes, images of which are extracted from the Sloan
Digital Sky Survey. Within Unreal Engine 4, galaxy images are
transformed into textures and dynamic materials (with appropriate
transparency) that are applied to static mesh objects with
appropriate sizes and locations. To ensure excellent performance,
these static meshes are ``instanced'’ to utilize the full capabilities
of a graphics processing unit. Additional components include a
dynamic system for representing accelerated-time active galactic
{nuclei}. The end result is a visually realistic large scale universe that can be
explored by a user in real time, with accurate large scale
structure. {Astera is not yet ready for public release, but we are exploring options to make different versions of the code available for both research and outreach applications.} }


\begin{document}


\section{Introduction}

The predominantly accepted cosmological paradigm, $\Lambda$CDM, predicts that structure in the universe forms via the collapse of dark matter into distinct filaments, voids and haloes \citep{ZelDovich1970, Peebles1980}. This underlying substructure acts as a tracer for baryonic matter, which falls into the potential wells of dark matter haloes and forms galaxies. It is therefore predominantly agreed that the large scale, 3-Dimensional structure of the universe is dictated by the structure of underlying dark matter. This structure is notoriously difficult to visualize in two dimensions, and since the early days of cosmological research, numerous attempts have been made to visualize the '`cosmic web''. Initial attempts by, for example, \cite{Doroshkevich1978} showed a representation of the cosmic web as a simple 2D '`slice'', with simple-plotted points representing the density distribution. Since then, increasing advances in digital graphics have allowed for significantly more visually impressive images---e.g., \cite{Springel2005} and \cite{Abel2012}. Despite the impressive quality of these images, it is clear that a 2-Dimensional projection of the large scale structure of the universe cannot capture the high complexity its full 3D structure, so more elaborate projections in pseudo-3D or utilizing iso-surface density have been developed---e.g., in \citep{Klypin1983}. The advent of computer graphical rendering has allowed for truly breathtaking video representations of the large scale universe in projected 3D, such as \cite{MilleniumVids}, \cite{BolshoiVid}, \cite{IlustrisVid}, and \cite{Flight}.

Digital graphics can be broadly categorized into ``pre-rendered'' graphics and ``real time'' graphics. The former requires pre-processing of the digital assets (often at great computational expense, but only once) into a series of frames that are assembled into a video format that is fixed in scope but can be replayed on virtually any device. This is most commonly seen in modern digital animation, commonly used in the entertainment industry. The latter, ``real time'' rendering, requires digital assets to be processed ''on the fly'', and must be processed quickly enough to ensure that the image can be recreated many times a second. This requires significant computational power, but crucially allows for user interactivity, with potentially unlimited scope for the user having dynamic control over the experience. Real-time rendering has, until recently, been restricted solely to video games, with visual quality notably worse that pre-rendered graphics. However, in recent years the quality of real time graphics has notably improved, allowing for their use in a wide variety of applications, such as on virtual film sets \citep{Mando}.

Real-time rendering depicting extragalactic scales have not yet been explored with full scientific accuracy. Several projects have come close; Celestia \citep{Celestia} is a free visualizer for many astronomical objects. Its open source nature has lead to numerous extensions, including planned cosmological visualizers. There are numerous sky visualizers that offer an extragalactic view of the universe based on astronomical imagery, such as WordWide Telescope \cite{WorldWide}, Google Sky \cite{GoogleSky} and others, focusing on real data. Universe Sandbox \citep{UniverseSandbox} is an interactive educational software application that simulates gravitational effects in various scales, including interactions between pairs of galaxies. SpaceEngine \cite{SpaceEngine} is a remarkable achievement, and is capable of procedurally generating a vast universe the user can explore with scales ranging from individual planets to the extragalactic.

Each of these projects, mostly focused on small-scale bodies such as planets and stars, is limited by the compelling need to simulate a vast range of length scales, understandably focusing on the solar system and similar bodies. This methodology inevitably produces a lack of accuracy when simulating the large scale structure of the Universe. In this paper, we present our solution: ``Astera'', a real-time cosmological visualization tool created using Unreal Engine 4.

This paper is structured as follows. In Section 2 {we} discuss the '`assets'' behind Astera, which pertains to the creation of the underlying example galaxy catalogue that forms the theoretical foundation of Astera, and the astronomical images used to represent galaxies. This is followed (Section 3) by a description of the technical implementation of Astera itself within Unreal Engine 4. In Section 4 we present the 3D universe that Astera creates, and we conclude in Section 5. 

\section{Assets}

Assets, as mentioned previously, are the reusable and replaceable components of a digital project. It is worth emphasizing that these components are able to be changed with a minimum of effort, so although integral parts of the experience they are not fixed parts of the software. This includes the underlying galaxy catalogue, and the astronomical images used to represent galaxies.

\subsection{Galaxy Catalog}

Mock galaxy catalogues are artificial datasets containing parameters for a large number of synthetic galaxies, extracted from simulations that utilize our best understanding of galaxy evolution. The comparison between mock catalogues and observational data is vital for probing underlying physical processes, but mock catalogues are also in demand for the calibration of the next generation of extra-galactic observations (e.g., Euclid, Athena). Mock catalogues are extracted from simulations that follow the evolution of galaxies over cosmic time, and from this a ``light cone'' is normally constructed that represents the evolutionary state of objects at varying redshifts. In the case of Astera, we are interested in (at least initially) creating a catalogue that can be ``explored'' by the user in real time, imitating non-physical superluminal speeds (or a universe ``frozen'' with no relative motions of galaxies, no cosmological expansion, etc). Therefore, the creation of a light cone is not necessary, and a simple cosmological volume will suffice.

{It should be noted that Astera itself is Cosmology independent, as it simply presents coordinates and galaxy imagery. The underlying mock catalogue can be of any cosmology desired, and in this case we adopt a standard flat $\Lambda$CDM paradigm with $H_0$ = 70, $\Omega_\Lambda$ = 0.7, $\Omega_0$ = 0.3. In this section, we describe a simple ``recipe'' that produces what the authors consider to be a reasonable mock catalogue for showcasing Astera, but in principle any catalogue could be used.}

As appropriate for the $\Lambda$CDM paradigm, the foundation of our catalogue is a Dark Matter N-Body simulation. We use both the Bolshoi \cite{Klypin2011} $(500 h^{-1} Mpc)^3$ and the Multi-Dark \citep{Klypin2016} $(1000 h^{-1} Mpc)^3$ simulation catalogues\footnote{Strictly speaking, these simulations adopt cosmologies that are slightly different to ours. However, for the purposes of this paper, they are sufficiently similar.}, depending on the volume desired (as larger volumes require more powerful hardware, although it should be noted that volumes can also be `'cropped'' to manage performance). The most important variables in this catalogue are naturally the 3D coordinates (X, Y, Z), as they dictate the 3D positions of the haloes in the virtual world, but also vitally important are the virial masses of the haloes, upon which we use statistical `'empirical'' relations to construct our model. We therefore assume that a galaxy exists at the centre of every dark matter halo resolved within the catalogue, and assign the stellar mass.

\begin{figure}[H]
\centering
\includegraphics[width=10 cm]{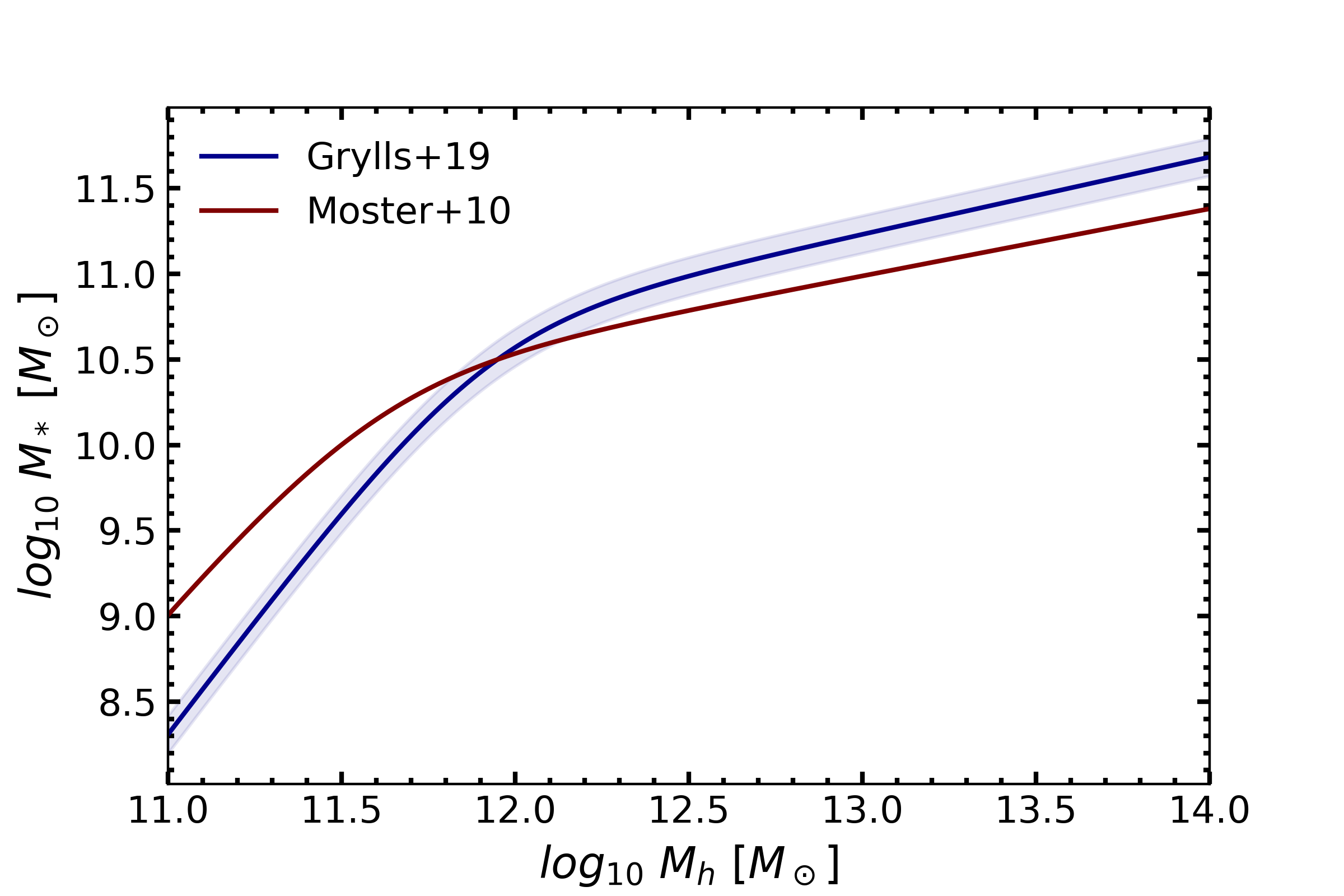}
\caption{Comparison of Halo Mass to Stellar Mass relations from \cite{Grylls2019STEEL} compared to \cite{Moster2010}.}
\label{fig:HMSM}
\end{figure}

The importance of robust and self consistent relations between underlying dark matter and galaxy stellar masses cannot be underestimated. As demonstrated by \cite{Grylls2019STEEL} and \cite{Grylls2019STEEL2}, even small variations in the Halo Mass-Stellar Mass (HMSM) relation can yield large (and often non-physical) variations in the satellite accretion rate, required star formation rate and pair fractions. While many mock catalogues will undoubtedly have established Stellar Masses, we adopt the \cite{Grylls2019STEEL} HMSM relation, a parametric stellar mass to halo mass relation, a variation of the relation presented in \cite{Moster2010}. For completeness, the full relation is shown in Equation (\ref{eq:GryllsHMSM}), with tabulated parameters in Table \ref{tab:HMSMtable}, and a plot of the relation is shown in Figure \ref{fig:HMSM}.

\begin{equation}
\begin{array}{l}
M_*(M_h, z) = 2M_hN(z) \left[\left( \frac{M_h}{M_n(z)}\right)^{-\beta(z)} + \left( \frac{M_h}{M_n(z)}\right)^{\gamma(z)} \right]^{-1}\\

\;\;N(z) = N_{0.1} + N_z \left(\frac{z - 0.1}{z + 1}\right)\\

\;\;M_n(z) = M_{n,\,0.1} + M_{n,\,z} \left(\frac{z - 0.1}{z + 1}\right)\\

\;\;\beta(z) = \beta_{0.1} + \beta_z \left(\frac{z - 0.1}{z + 1}\right)\\

\;\;\gamma(z) = \gamma_{0.1} + \gamma_z \left(\frac{z - 0.1}{z + 1}\right)
\end{array}
\label{eq:GryllsHMSM}
\end{equation}

\begin{table} [h]
	\centering
	\caption{Parameters for equation \ref{eq:GryllsHMSM}.}
	\label{tab:HMSMtable}
	\begin{tabular}{lccccr}
		\hline
		$ $ & $M_n$ & $N$ & $\beta$ & $\lambda$ & $\sigma$\\
		\hline
		Central, $z=0.1$ & 11.95 & 0.032 & 1.61 & 0.54 & 0.11\\
		Total, $z=0.1$  & 11.89 & 0.031 & 1.77 & 0.52 & 0.10\\
		\hline
		Evolution, $z > 0.1$ & 0.4 & -0.02 & -0.6 & -0.1 & N/A\\
		\hline
	\end{tabular}
\end{table}

\noindent {where $M_*$ represents the stellar mass of the galaxy $M_h$ represents the host halo mass. The associated parameters $N$, $M_n$, $\beta$ and $\gamma$ are values set to describe the relation at varying reshifts. In table \ref{tab:HMSMtable}, the parameters are shown for central and satellite galaxies at $z=0.1$ and evolving values (the subscript $0.1$ referring to the value at $z=0.1$, and the subscript $z$ referring to the evolving value). This equation and table are reproduced from \cite{Grylls2019STEEL}, wherein more details can be found.} 

\begin{figure*}[t!]
    \centering
    \begin{subfigure}[t]{0.49\textwidth}
        \centering
        \includegraphics[width=\textwidth]{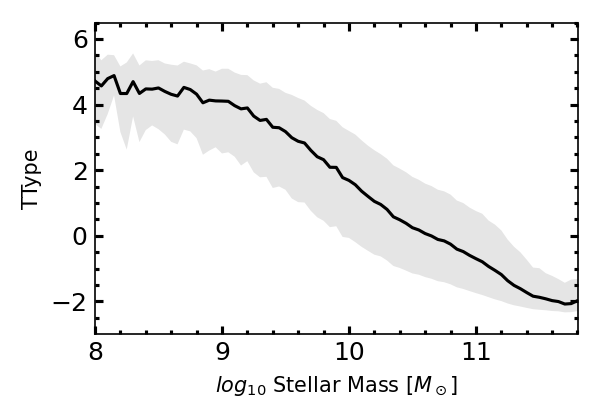}
        \caption{TType}
    \end{subfigure}%
    ~ 
    \begin{subfigure}[t]{0.5\textwidth}
        \centering
        \includegraphics[width=\textwidth]{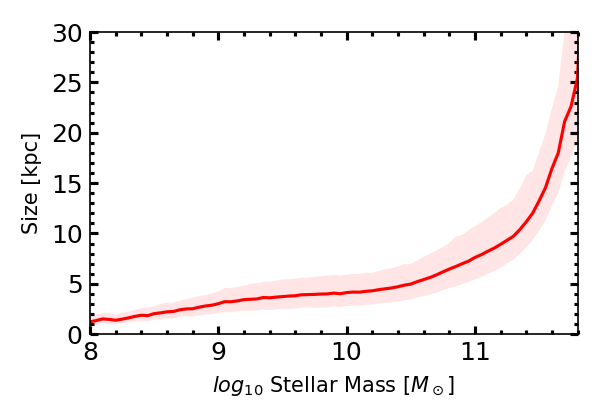}
        \caption{Size}
    \end{subfigure}
    \caption{(\textbf{a}) Morphological type (TType) and (\textbf{b}) Size vs Stellar Mass relations within the SDSS, used to assign the respective parameters to the catalogue. The shaded regions show the $1\sigma$ uncertainty in these parameters.}
    \label{fig:SDSS}
\end{figure*}

Assigning stellar mass allows for a robust foundation upon which further properties can be built. Aside from 3D distribution, the most obvious property required for a visually realistic galaxy catalogue is obviously morphological classification. This was assigned using phenomenological relations derived from the Sloan Digital Sky Survey (SDSS) data \cite{SDSS2017}. Binning the SDSS by stellar mass allows for an approximate Stellar Mass-morphological type (TType) relation (see Figure \ref{fig:SDSS} (\textbf{a})). {In short, the catalogue of galaxies is binned in appropriately sized bins of stellar mass, and the mean TType and scatter for each bin are recorded, producing an (average) relation allowing transformation from stellar mass to TType. With the associated scatter,} this relation can be applied to the simulation catalogue for a statistically comparable distribution of morphological types to the real universe. Each galaxy is therefore assigned a TType representing its morphological classification based on its stellar mass. A caveat here is that this distribution is only valid at low redshifts, limiting this technique to catalogues representing the universe in the ``present day''. This limitation is not insurmountable, as numerous semi-analytic models and semi-empirical models are capable of predicting morphological abundances at varying redshifts, and their datasets could be easily be used in Astera. 

A galaxy's physical size (parametrised by the effective radius $R_{\mathrm{eff}}$) is also naturally important for visual realism. These were assigned in identical fashion to morphological types, using the mean relations from the SDSS (again see Figure \ref{fig:SDSS} (\textbf{b})).

\begin{figure}[H]
\centering
\includegraphics[width=0.7\textwidth]{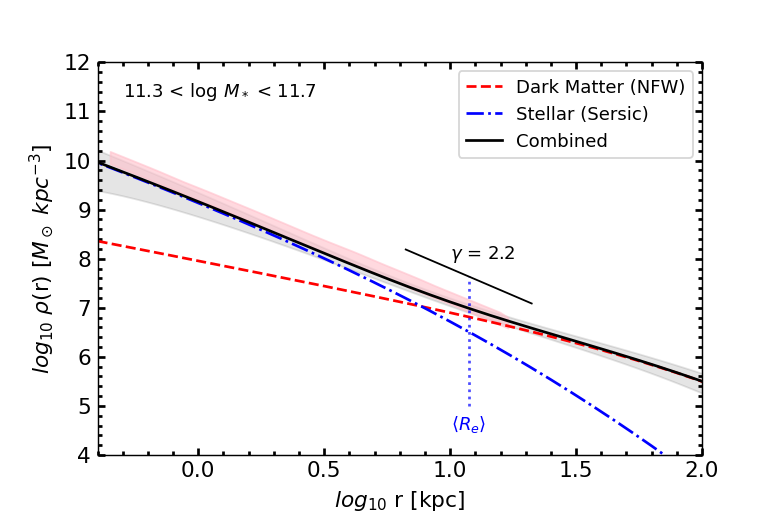}
\caption{{Predicted (average) 3D density profile haloes hosting elliptical galaxies of mass $11.3 < \log_{10} M_*/M_\odot <11.7$ for the sample catalogue. The solid black line represents the combined density, whilst the blue (dot-dashed) and red (dashed) lines represents the average stellar (S\'{e}rsic) and dark matter (NFW) components, respectively. The pink shaded region represents the empirical fit from \cite{Cappellari2015}. The grey shaded region shows the $1\sigma$ dispersion of the total density. The vertical blue dotted line shows the average half light radius, and the solid short black line shows (for comparison) the established slope of $\gamma = 2.2$ at this radius.}}
\label{fig:MassProfile}
\end{figure}

{A simple validation to show that these components are ``working together'' as expected is shown in Figure \ref{fig:MassProfile}, where we show the average predicted 3D mass density profile for elliptical galaxies of stellar mass $11.3 < \log_{10} M_*/M_\odot <11.7$. Each elliptical galaxy in this mass range is additionally assigned a S\'{e}rsic index (again according to the mean relation in the SDSS), a de-projected stellar mass density profile according to the prescription of \cite{Prugniel1997}, and an NFW profile \citep{Navarro_1996} with halo concentrations according to the model of \cite{Diemer2019}. Finally, the average mass density per bin of radius is calculated for the sample. This result is compared to the pure power law model of \cite{Cappellari2015} $\rho(r) \propto r^{-\gamma}$, where $\gamma = 2.2$ at $r \sim R_e$, derived from 2D stellar kinematics and strong lensing measurements. Specifically, \cite{Cappellari2015} inferred that $\langle\gamma\rangle = 2.19 \pm 0.03$, valid in the range $0.1Re$ to $4Re$ and $10.2 < \log_{10} M_*/M_\odot <11.7$. This result is shown as the pink stripe in Figure \ref{fig:MassProfile}, showing good agreement with the model at $\langle R_e\rangle$. The slopes at low radii were found to be very sensitive to the definition of $R_e$, in this case showing a slight divergence from the data.}

Finally, active galaxies were also considered. These were assigned using a bespoke methodology that will be fully described in Marsden et al. (in preparation) and Allevato et al. (in preparation), but in general the central supermassive black hole mass was assigned from the stellar mass using the unbias relation from \cite{Shankar2016}, with appropriate Eddington ratios and X-Ray luminosity assigned using a Schechter function with values chosen to fit the known X-Ray luminosity function and Eddington ratio distribution. The ``Duty Cycle'' $U$ or associated probability of a Supermassive Black Hole being active, was for now set to $U$$\sim0.1$, although a more complete treatment of this process will be described in Marsden et al' (in preparation and Allevato et al. (in preparation)).

\subsection{Galaxy Imagery}

An obviously vital part of Astera is the galaxies. We elected to use actual astronomical imagery, and in section 3 we discuss this choice further. For now, we  discuss the images themselves. Getting both good quality and a large variation in galaxy images will contribute greatly to the user experience, so various approaches were considered.  There are relatively few high resolution images from the Hubble Space Telescope (HST), but much more diverse but lower resolution images from the Sloan Digital Sky Survey (SDSS), which may be more appropriate. Further work was also put into investigating the feasibility of creating our own artificial images based on a relatively small number of starting parameters; this is known as ``procedural generation''. This has been done before using Neural Networks and Hydrodynamical simulations \citep{nguyen2016}, but a considerably simpler (at least initially) approach was desired here. We therefore focused on acquiring actual astronomical imagery.

The actual astronomical imagery from the SDSS is available on their website. Properly identifying likely target galaxies is challenging. A list of targets, identified by \cite{meert2014catalogue}, was processed into a list of likely candidates. The most promising targets had a large angular diameter distance, allowing the most high quality galaxies to be quickly identified. Creating a visually appealing astronomical image from the data acquired in the appropriate bands is a somewhat subjective process. Of vital importance is the ``stretch function'' applied to the data, which applies a mathematical transformation to the pixel values to (ideally) enhance brighter areas and saturate darker areas, eliminating noise. In this project, the stretch function was applied using the ``Fits Liberator'' \cite{Fitsliberator} software, which offers a GUI to allow the user to select and tweak the stretch function parameters on the fly. The fits Liberator can also export the image into the GIMP \citep{GIMP} image manipulation software as Red, Green and Blue (RGB) components, creating the image. Some additional tweaking was required at this point. Background stars and galaxies have to be carefully erased (GIMP offers many tools to do this, the best of which are actually designed to interpolate out blemishes from images of human skin, but equally applicable to ``blemishes'' in the sky), and cubic interpolation to re-scale all the images to the same size. 

\begin{figure}[H]
\centering
\includegraphics[width=\textwidth]{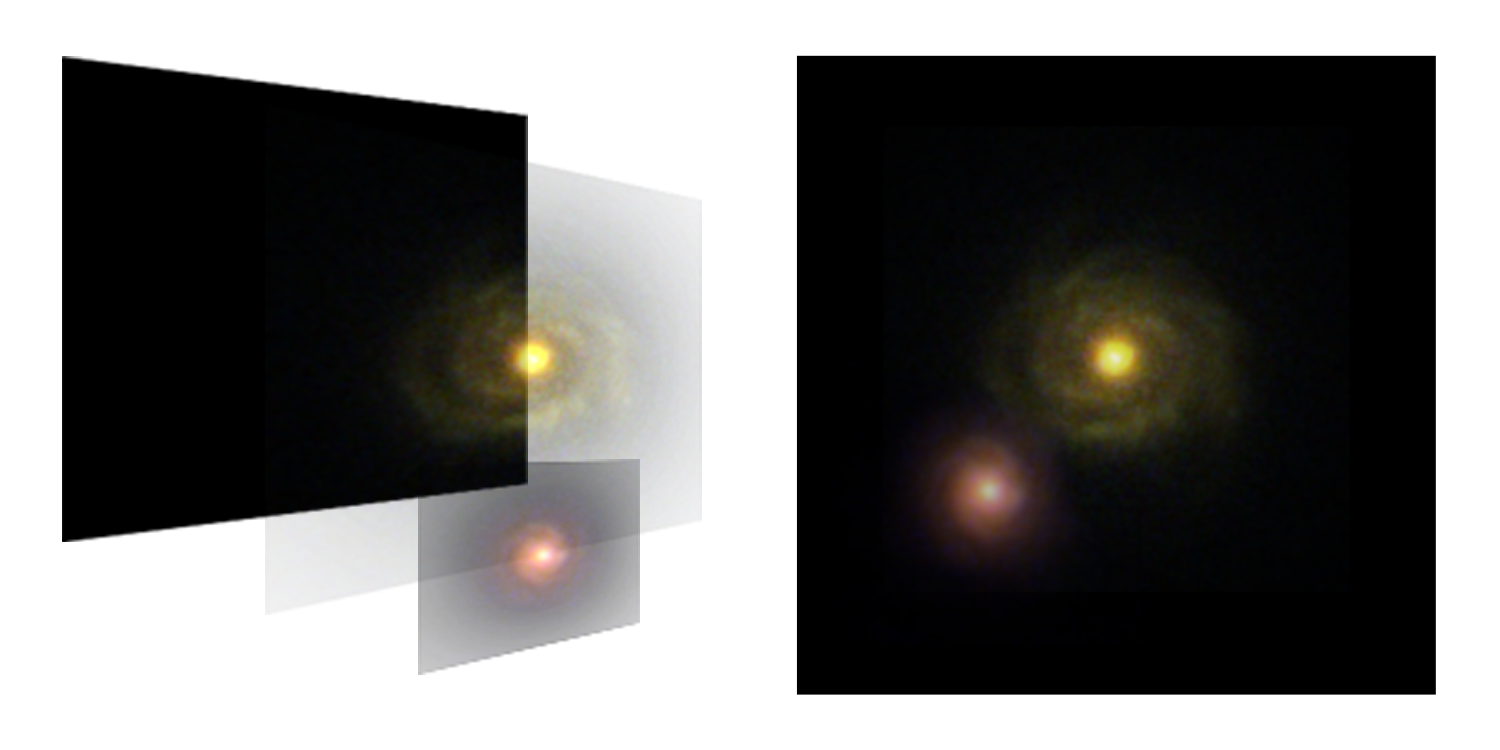}
\caption{Example depiction of the importance of the alpha channel. When projected onto each other, the alpha channel allows the galaxies to appear as diffuse objects, hiding the sharp edges.}
\label{fig:alpha}
\end{figure}

Finally, one of the most important steps is the construction of an ``alpha'' channel. This image layer contains values that dictate the transparency of the corresponding pixel in the image; an alpha value of one would signify a fully opaque pixel, whereas an alpha value of zero would be fully transparent. Setting this up properly is vital for Astera, as the transparency is the property that will ``soften'' a geometric mesh into a believable diffuse galaxy. The alpha channel was assigned in this case using the sum of the RGB layers, appropriately normalized ``by eye'' to appear visually realistic. The importance of this channel is demonstrated in Figure \ref{fig:alpha}.

\begin{figure}[H]
\centering
\includegraphics[width=\textwidth]{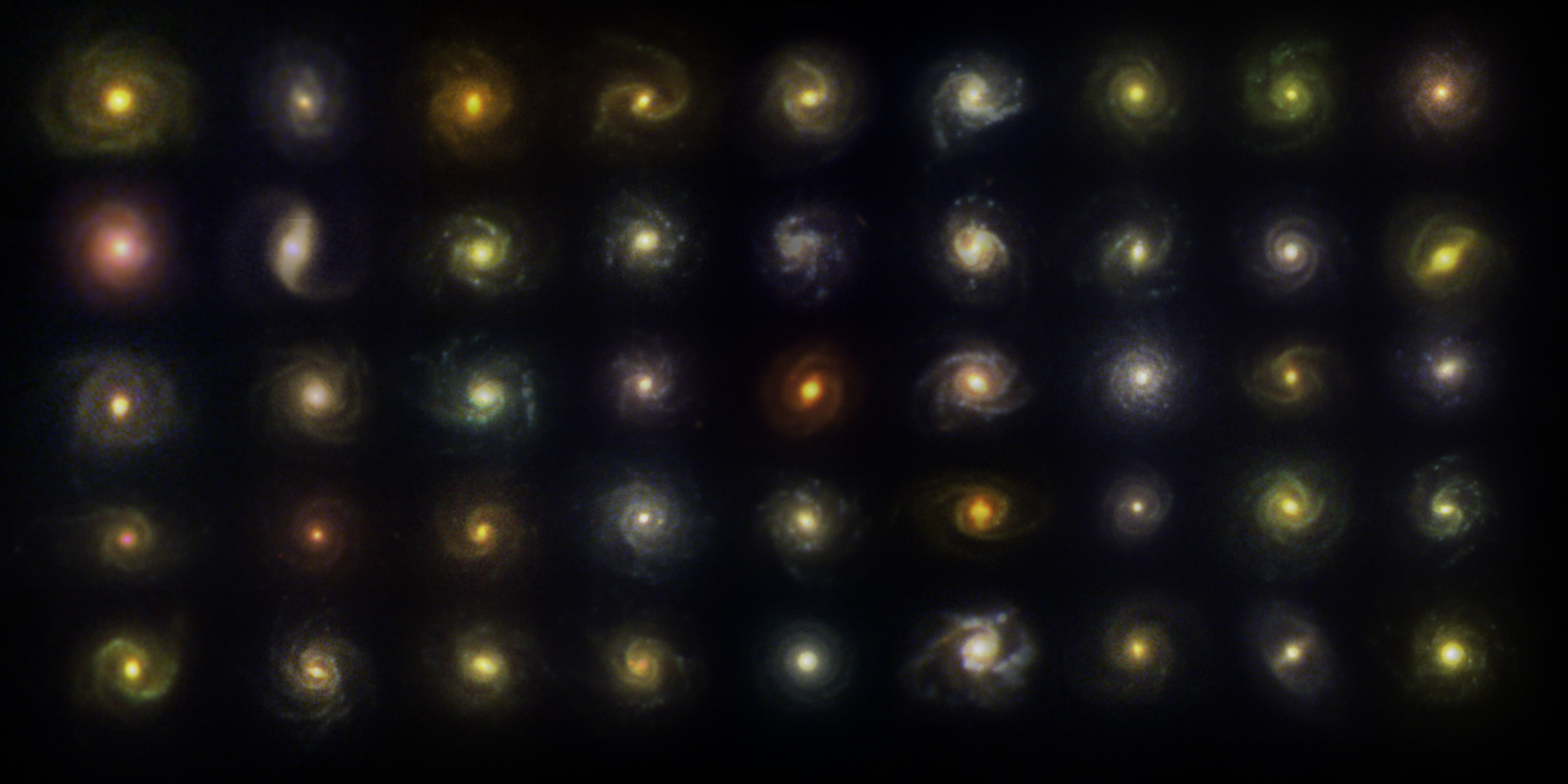}
\caption{Composite images of 45 spiral galaxies extracted from the SDSS, processed to be visually pleasing.}
\label{fig:gals}
\end{figure}

Using this method, a few hundred distinct galaxies were extracted and processed from the SDSS dataset. Some examples are shown in Figure \ref{fig:gals}. While these galaxies are not as high resolution as HST images, their diversity allows for a wide range of galaxy types and morphological classifications. The Stellar Mass and Morphological types of each galaxy was also recorded for later use.

\section{Unreal Engine}

The Unreal Engine is a game creation engine. Technically, a complete suite of creation tools, Unreal is best known as a package containing a rendering engine, sound engine, physics engine, gameplay framework, animation, artificial intelligence, networking, memory management and parallel processing support. These reusable software components act like a vast library of tools that can be utilized by the game developer to assemble their game. Strictly speaking, references to the Unreal Engine in this paper indicate Unreal Engine 4 (sometimes referred to as ``UE4''), the fourth release of the software. Unreal Engine 4 significantly overhauled many of the features of Unreal Engine 3 when it was released in 2014, so many of the tools and features discussed and utilized as part of this project may not be available in earlier versions of the software. Unreal is freely available for non-commercial use.

There are various possible approaches to render galaxies in the Unreal Engine. It is possible to render Galaxies in real time as a system of diffuse particles. These systems are often both computationally expensive and visually unrealistic, so a different approach was considered in the development of Astera. Because each galaxy will be relatively small on the scales that we are interested in viewing, a single geometric object with an applied material (sampling a texture based on actual astronomical imagery) will suffice, and free up resources to show a greater quantity of galaxies in the game world. 

\begin{figure}[H]
\centering
\includegraphics[width=\textwidth]{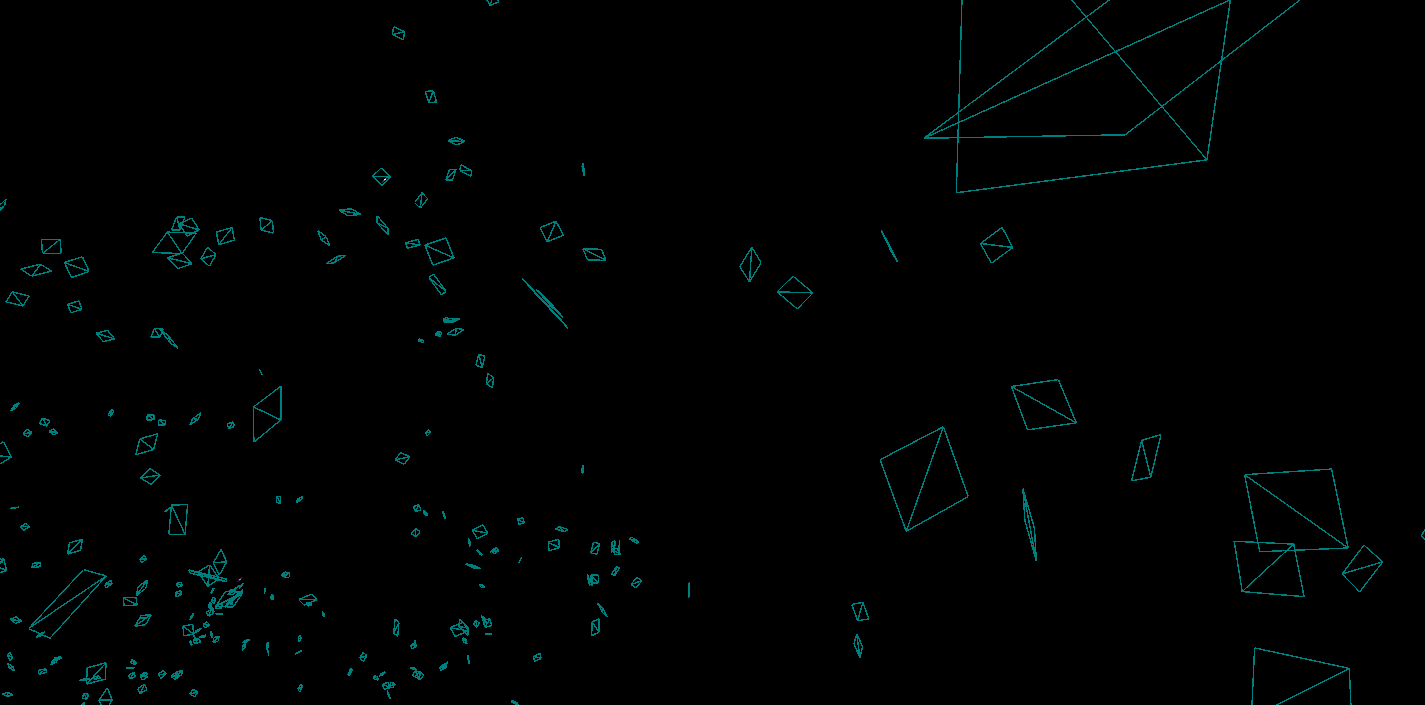}
\caption{A wireframe view of a small area with Astera. The instanced static mesh objects used in Astera to represent galaxies can be seen.}
\label{fig:wireframe}
\end{figure}

This is done as follows. Spiral Galaxies are generally disc-shaped, and Ellipticals are vaguely spherical. Although it is possible to construct 3D versions of these shapes, a more geometrically complex shape is both more expensive to render in bulk and harder to properly configure with a material. Galaxies are diffuse objects, but a simple approximation allows for a ``Static Mesh'' object that defines the geometric shape of the galaxy to be used as a visual proxy. A static mesh will have a ``material'' applied to it, which can contain colours and textures (in this case, the galaxy image), but also a large amount of additional complexity (such as transparency or varying textures in time, used for AGN activity). Based on this, every galaxy in Astera is based on a simple two polygon plane to keep the geometry simple (see Figure \ref{fig:wireframe}, with additional ``work'' being done by the material applied to it). As the galaxies will be kept relatively small with respect to the camera, this should not be problematic.

This is a fair approximation for spiral galaxies. On the other hand, Elliptical galaxies are spheroids, so must be represented differently. The obvious choice would, therefore, be a spheroidal mesh. However, this does not correctly represent the diffuse nature of an Elliptical. A far more realistic choice requires some visual trickery; if the elliptical mesh always appears ``face on'', then a moving camera will always perceive the object as a diffuse sphere. This requires all elliptical galaxies to (for now) be essentially spherical, but this is an acceptable limitation. It may be possible in the future to dynamically change the scaling of the mesh to a greater extent on one axis based on the camera's position, thereby enabling more irregular ellipticals. Lenticular galaxies are problematic (as they require a combination of these effects) and are therefore structurally treated as spirals in the first version of Astera. Unreal Engine's material system is powerful enough to allow this modification to take place within the material itself.

Rendering a large number of objects inevitably leads to a performance penalty. It should be noted at this point that performance in real time rendering, although measurable in many ways, is generally parametrised using ``Frames Per Second'' (FPS), a number representing the mean number of updates to the screen per second (higher is better). Values between $30$--$60$ FPS are generally considered acceptable. Naturally, a more complex scene requires more computational resources, and may therefore result in a lower FPS. 

Extraordinary care must be therefore taken when rendering large numbers of distinct mesh objects in Unreal. The best approach in this case requires careful thought. Graphical Processing Units (GPUs) are very efficient at drawing polygons, but need to be ``fed'' these data by the (comparatively slow) CPU. Whenever a new object is ``drawn'' to the screen, it requires a separate CPU call (going through the graphics driver). An alternative is to combine these objects into a single mesh; requiring only one draw call. The only disadvantage to this approach is it requires all objects to have the same material and textures. As we are ``duplicating'' our galaxies anyway, this is not a problem; we simply create one object per available galaxy image asset. Although we are performing more draw calls, the overall number of CPU calls is still significantly lower (hundreds as opposed to millions), so the performance is still excellent. A mesh created in this way is called an ``instanced'' static mesh (ISM), and is the technology that makes a universe as large as Astera's possible.

Based on this technology, spawning the objects takes place as follows (this process is implemented in C++, using Unreal Engine's Application Programming Interface). For each galaxy image extracted for the SDSS, a parent unreal actor (with no mesh of its own) is spawned and the image and the properties of the imaged galaxy are assigned to it. Next, each galaxy in the catalogue is assigned to the object with its ``nearest'' properties in TType-Stellar Mass space. Practically, this means that the galaxy is assigned to the object that minimizes $A$ in the following relation:

\begin{equation}
    A^2 = \alpha \Delta T_{type}^2 + \beta\Delta M_{*}^2
\end{equation}

\noindent {where} $\Delta T_{type}$ is the difference in TType between the galaxy and the object, $\Delta M_*$ is the difference in stellar mass. The constants $\alpha$ and $\beta$ are adjustable normalization factors to ensure the relative weighting is appropriate. If $\Delta M_{*}$ was in units of $\log_{10} M_{\odot}$, then $\alpha=\beta=1$ was found to be a reasonable choice. The end result is that there will, of course, be many duplicates; individually assigning unique galaxy images to every galaxy in the catalogue is computationally unfeasible, and the aforementioned ISM technology requires duplicates. The aim here is to have a sufficient number of unique galaxies so that duplicates are not noticeable to the user, but sufficiently few to preserve performance. An appropriate compromise was a few hundred unique galaxy images, compared to the millions of galaxies in the catalogue.

Next, each object spawns its galaxies, each represented by a mesh with its galaxy image applied as a material. Spiral galaxies are assigned random orientations. At this point, different features can be activated, such as the technology to make the mesh follow the camera, approximating an elliptical. All galaxies have the capability to be AGN, where a central oscillating bright source was added with an active period equal to its duty cycle. The light curve of an AGN is not yet well constrained---e.g., in  \citep{Hopkins2009}---so the shape of the curve is approximated by the peak of a sine wave to give smooth variation in brightness. Naturally the period of this oscillation requires some level of approximation as to the `'rate of time'' at which the user perceives the universe, but keeping this period reasonably long ($\sim$ $60$ seconds) presented a visually pleasing result. Note that this is not entirely realistic, as on the timescales of AGN phases galaxies will themselves have moved and evolved. This is not something that Astera (yet) considers.

\section{Results}

\begin{figure}[H]
\centering
\includegraphics[width=\textwidth]{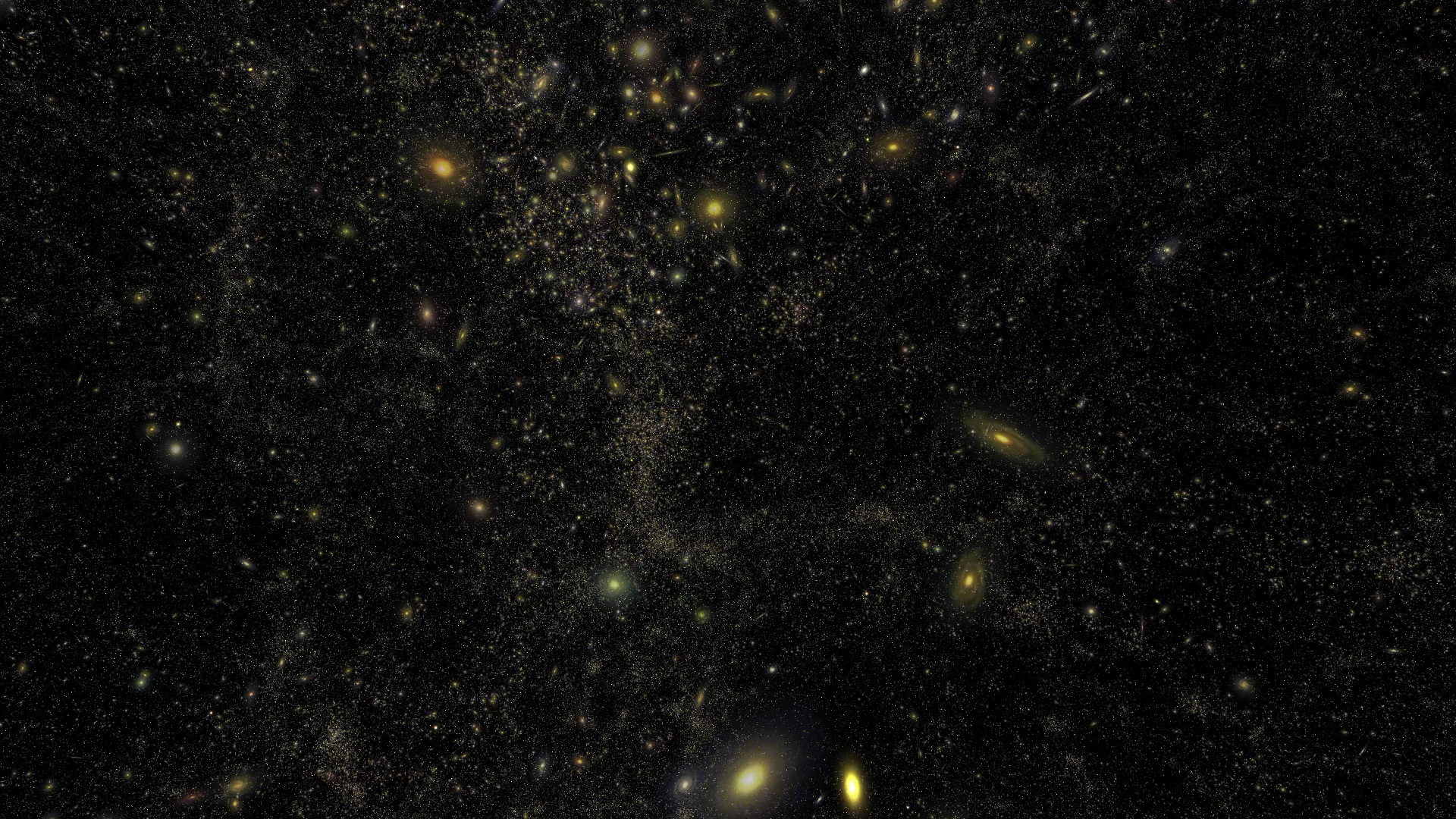}
\caption{A screenshot from Astera, showing some nearby galaxies in the foreground and a dense cluster/filament in the background.}
\label{fig:SS1}
\end{figure}

\begin{figure}[H]
\centering
\includegraphics[width=\textwidth]{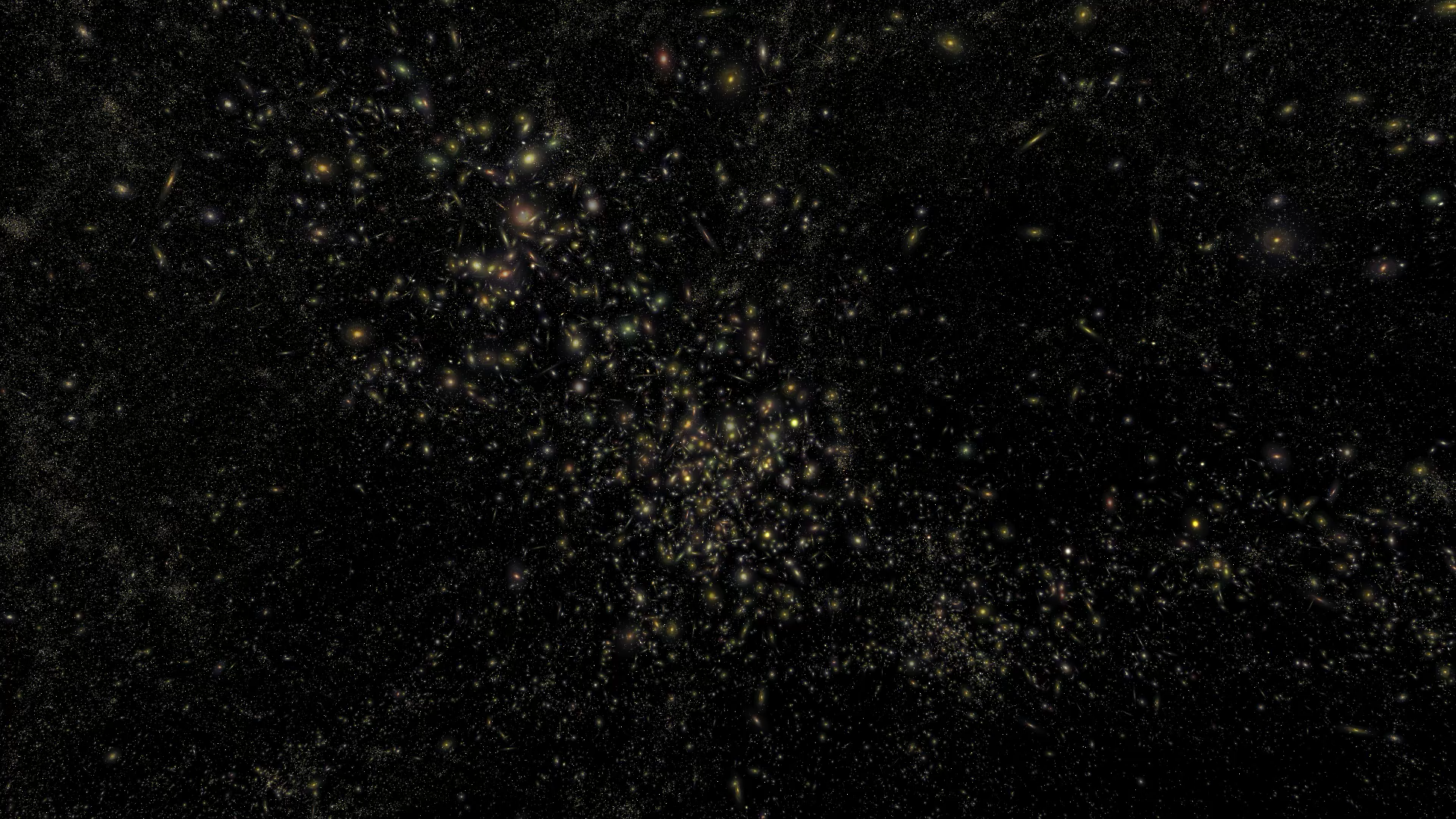}
\caption{A screenshot of Astera showing a relatively dense region. The structure of the cosmic web is just visible.}
\label{fig:SS2}
\end{figure}

\begin{figure}[H]
\centering
\includegraphics[width=\textwidth]{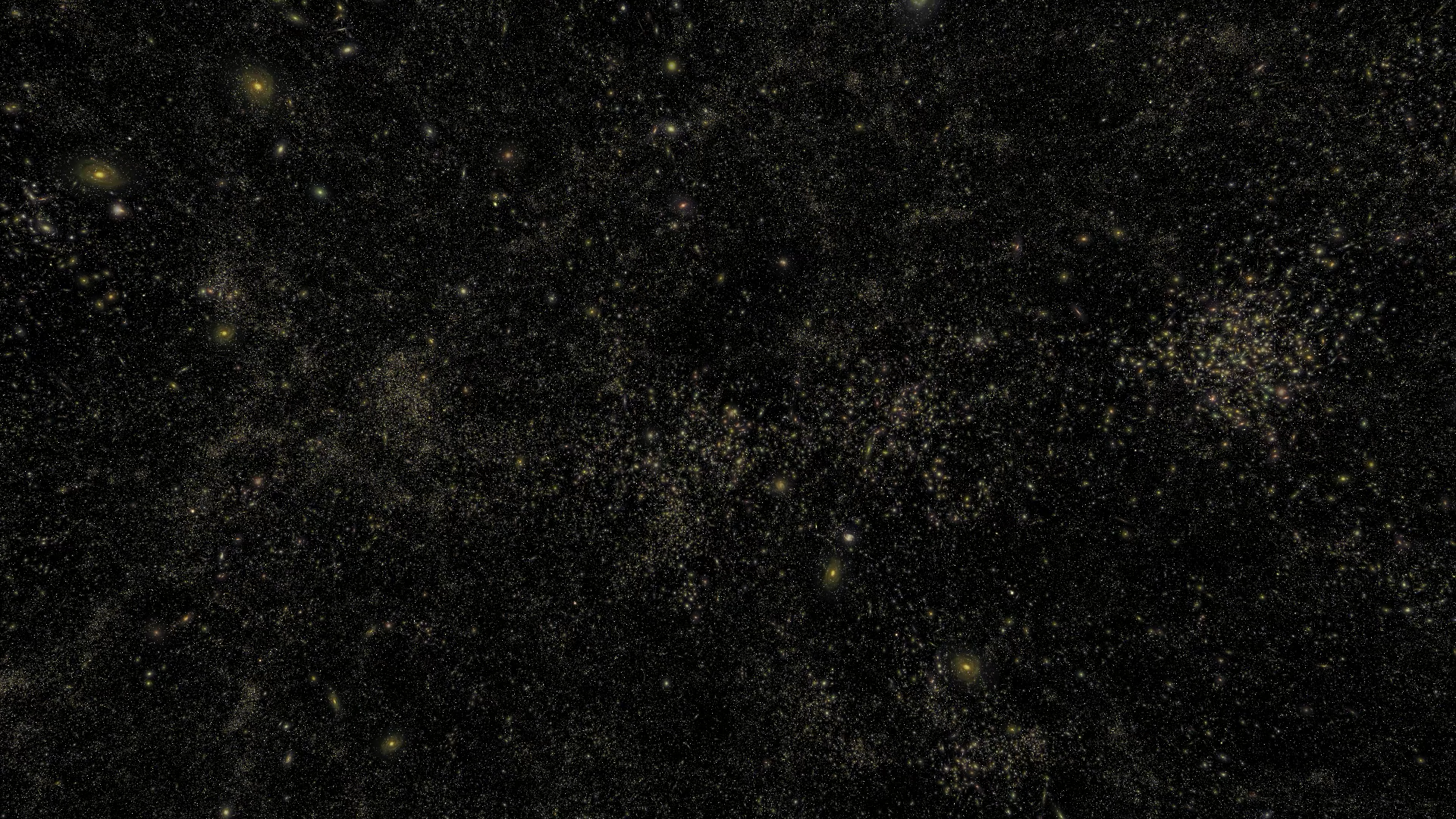}
\caption{A screenshot showing a distant view of four clusters joined by a filament.}
\label{fig:SS3}
\end{figure}

\begin{figure}[H]
\centering
\includegraphics[width=\textwidth]{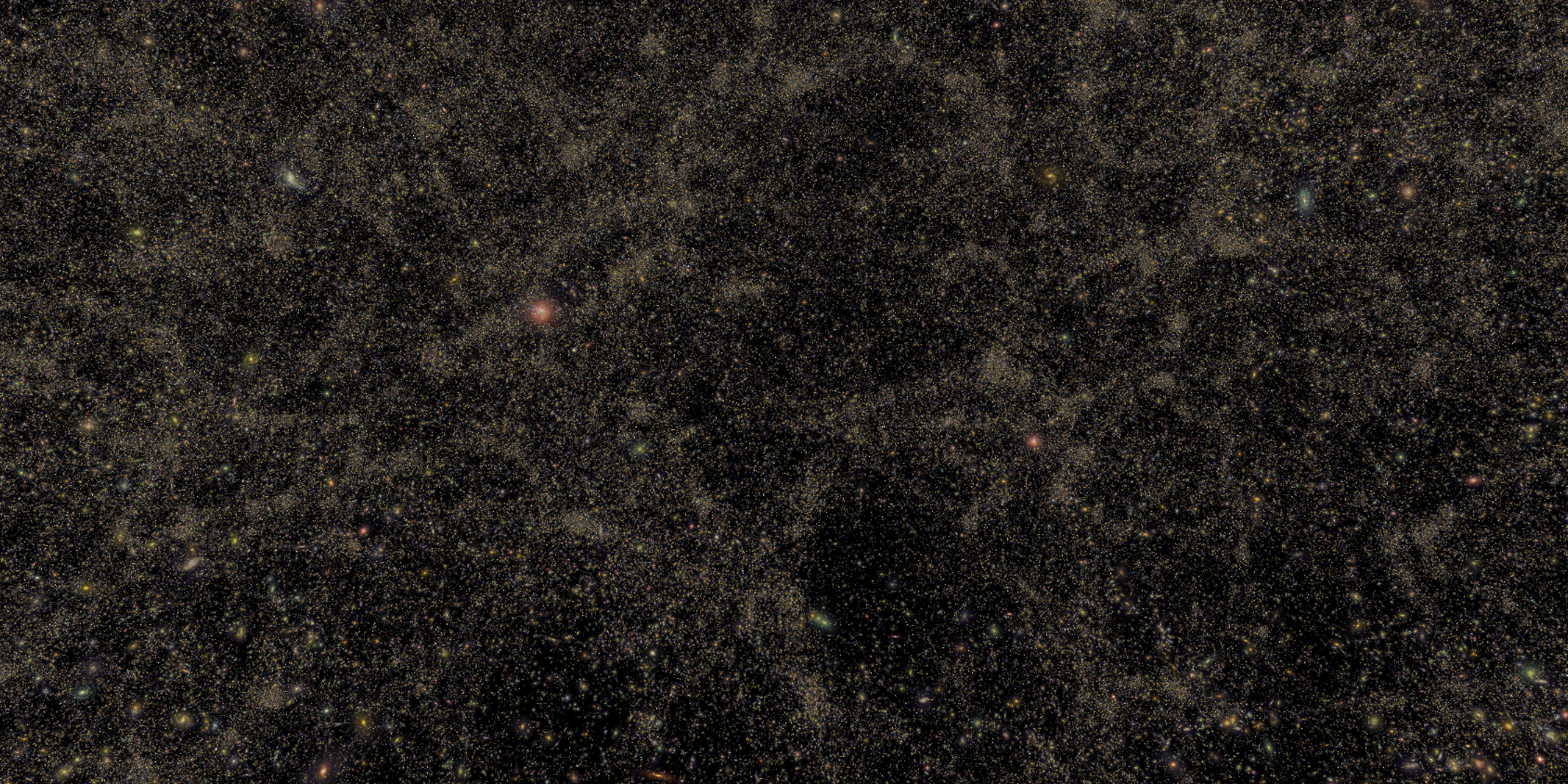}
\caption{A large scale screenshot showing many millions of galaxies within Astera.}
\label{fig:view}
\end{figure}

\begin{figure}[H]
\centering
\includegraphics[width=\textwidth]{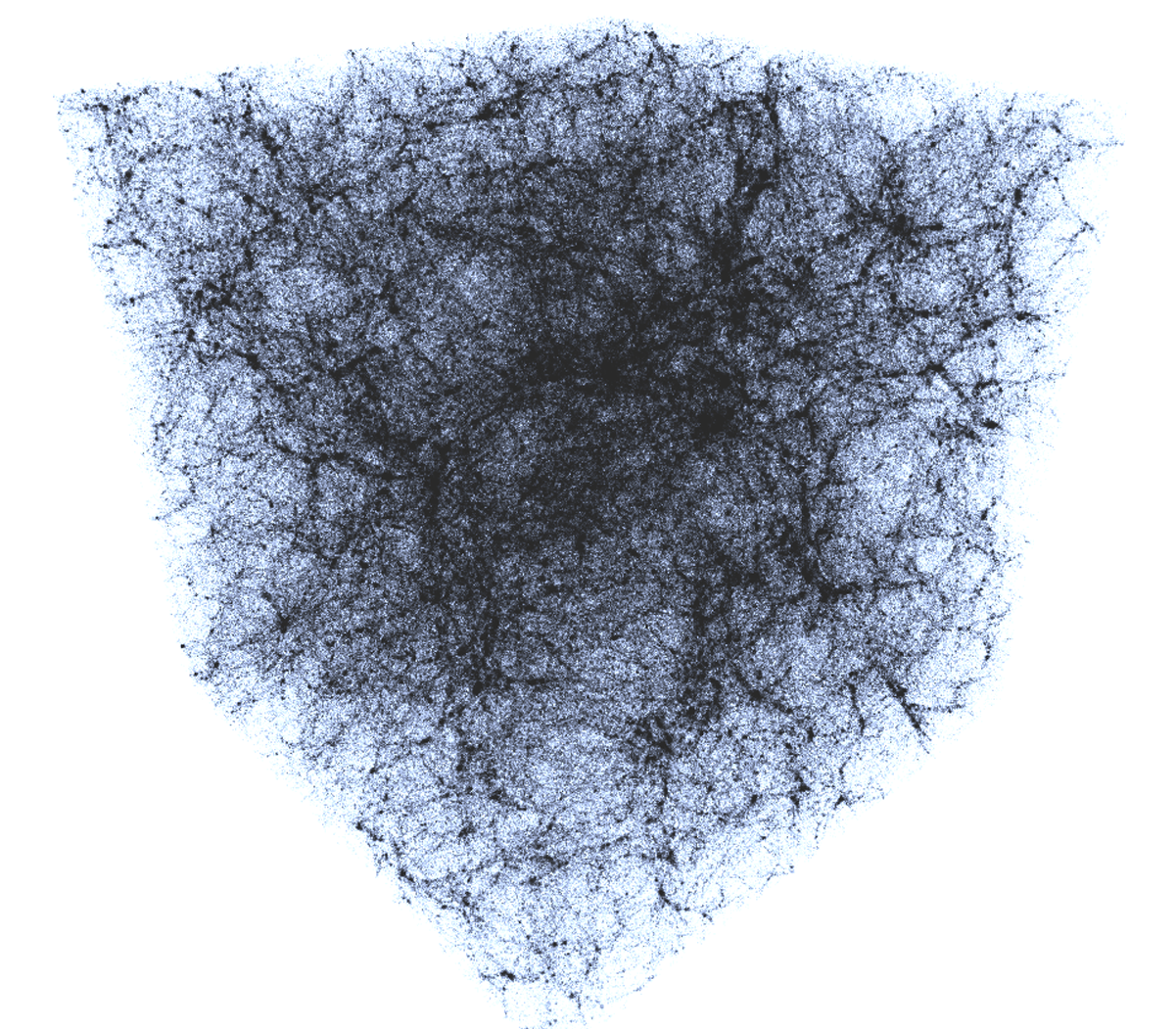}
\caption{A colour-inverted view of the full simulation volume. The large scale cosmic web is clearly visible}
\label{fig:box}
\end{figure}

Screenshots from Astera are shown in Figures \ref{fig:SS1}, 
\ref{fig:SS2}, \ref{fig:SS3}, \ref{fig:view} and \ref{fig:box}. Note that, in these images, galaxy brightness has been enhanced to improve visual clarity. Nonetheless, the experience Astera offers is hard to communicate in a document such as this. We strongly encourage the reader to watch the video available at \url{https://astera.soton.ac.uk/AsteraVid.mp4}\footnote{Due to the difficulty video streaming compression algorithms have with many small moving objects, it is recommended the reader download this video and play it locally.}, to get a full experience of what Astera can offer.

Astera is capable of displaying the entire galaxy catalogue with volume(1000 h $^{-1}$ Mpc)$^3$ at 60 FPS (on a NVidia Titan GPU), with each galaxy represented by a static mesh containing a galaxy image, effectively creating a `'universe'' for the user to explore. The player has control of the camera facing, position and movement in real time. The finite nature of the catalogue means that a user can exit the cube and view it from the '`outside'' (see Figure \ref{fig:box}).

Astera has also received preliminary adaption for Virtual Reality {using the Unreal Engine virtual reality tools, which ports Astera from projected 2D to simulated 3D when viewed through a VR headset. This is an entirely visual effect, but allows for some level of binocular vision,} enabling the 3D large scale structure of the universe to be experienced in {binocular} 3D for the first time.

\section{Discussion and Conclusions}

This project's objective was to create a visualization of a mock galaxy catalogue, rendered in real time. This would one of the first attempts to execute a fully realized real time rendering of the large scale universe. To power this project, the Unreal Engine 4 game engine was selected. A framework was constructed using C++ to read the data, create mesh instances and place the galaxies. The primary problem behind this task was performance; how many static mesh objects (representing galaxies) could be drawn to the screen at an acceptable frame rate. Remarkably, through the use of Unreal Engine's Instanced Static Mesh Technology, this was achieved to the extent that every galaxy in a full frame of a catalogue representing a (1000 h$^{-1} $Mpc)$^3$ box could be shown simultaneously on hardware running a NVidia Titan GPU at a full 60 FPS. Less powerful hardware can still run large volumes, with a NVidia GTX 760 rendering a (300 h$^{-1}$ Mpc)$^3$ box at 60 FPS.

\begin{figure}[H]
\centering
\includegraphics[width=\textwidth]{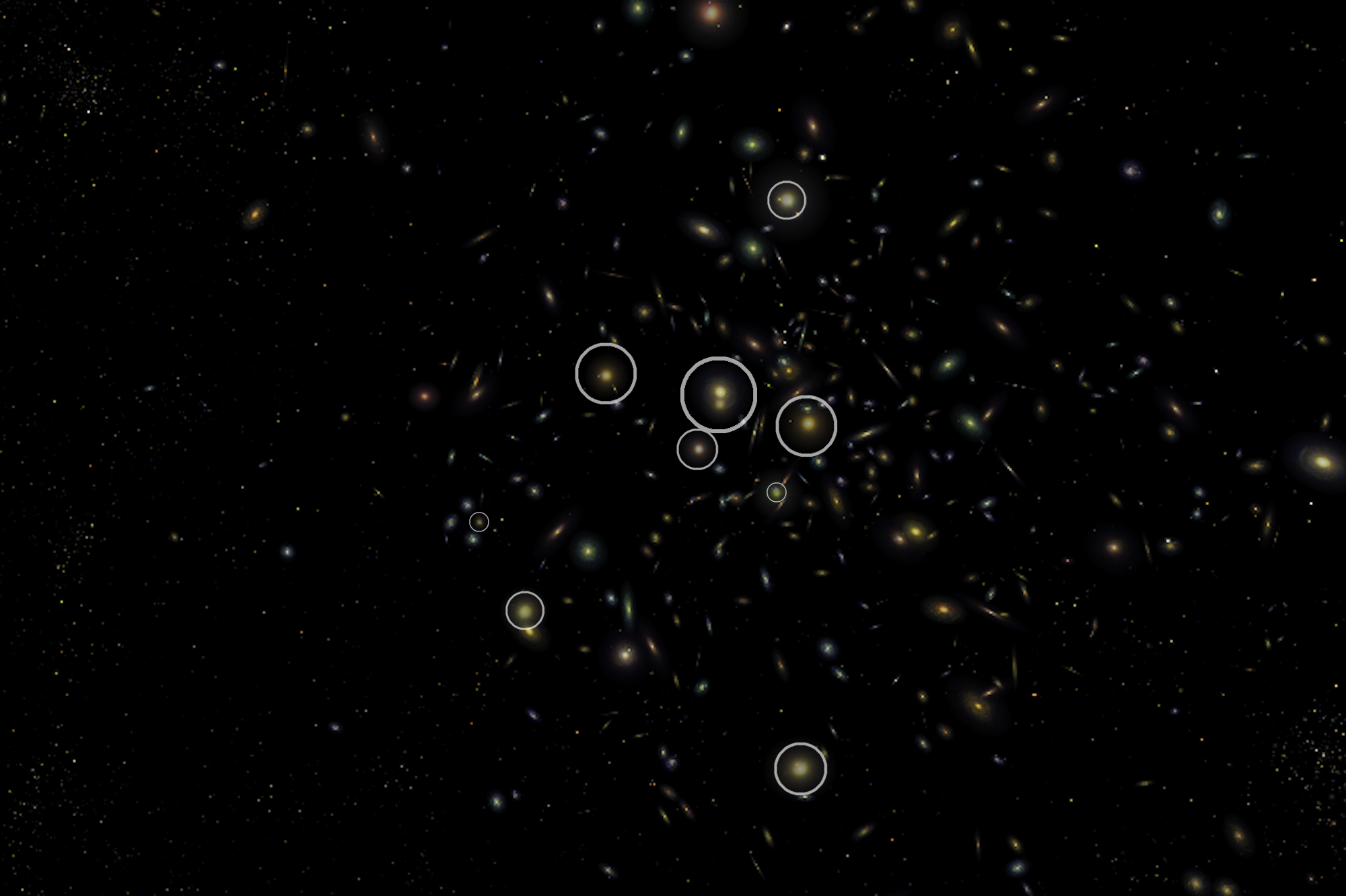}
\caption{A cluster of galaxies within Astera, where the large elliptical galaxies have been circled. The elliptical galaxies preferentially occupy the centre of the cluster, in line with observations.}
\label{fig:labeled}
\end{figure}

This was attained through intelligent use of Unreal's powerful and efficient Instanced Static Mesh Technology. Another success of Astera is its compatibility with any dataset. Any galaxy catalogue could easily be imported (within hardware constraints), a feature that could allow a researcher to replace the default data and explore their own universe. Exploring the universe within Astera reveals the large scale cosmic structure in a way that is vastly easier to understand than a 2D image or even potentially a video. The author noted that several '`wall'' like structures become visible in the cosmic web, which are not apparent in 2D imagery, in agreement with \cite{Diemer2017}. {The distribution of galaxy mythologies also seem to follow established trends (see Figure \ref{fig:labeled}), such as elliptical galaxies occupying the central regions of clusters---e.g., in \citep{Lucia2011}.} 

Astera is not yet ready for public release. The authors are exploring options for the most appropriate way to release Astera to astronomers and non-astronomers alike. Astera's ``scope'' is also relatively narrow, so as a relatively easy to use piece of software, Astera can be understood by the non-astronomer reasonably quickly, a feature that many similar Astronomical Visualizers do not have. Astera will soon be exhibited at a local attraction in Southampton at the Winchester Science Centre. This will feature a modified version of Astera, ``gamified'' to allow the user to scan and classify galaxies to gain points, with the aim of attracting a new generation to extra-galactic astronomy. 

{From an academic perspective, there are unique advantages to creating a mock universe that is rendered in real time. The first is simply the visual examination of Mock Galaxy catalogues (e.g., in preparation for large surveys); while not necessarily statistically robust, a quick visual examination can often reveal issues that would not otherwise be easy to detect. A more nuanced approach could be to reproduce the effects of a program, such as SkyMaker \citep{Bertin2009Skymaker}, where a mock galaxy catalogue is used to create a simulated astronomical image. Crucially, Astera would render this in real time, allowing for alternative images to be rapidly explored, or even a simulated sequence of images over a period of time, which could simulate data from time-variable objects, such as AGN.}

Future ``features' that the authors are exploring include:

\begin{itemize}
    \item {Dark Matter Viewer}. The conspicuous absence of Dark Matter in Astera would be remedied by a view mode that would show the dark matter substructure.
    \item {Time Evolution}. An exciting option which would essentially integrate Astera with a semi-analytic model, the motions and evolution of galaxies would be visible in (accelerated) real time. The user would be able to, at the press of a button, watch the universe evolve in front of them. This would dramatically increase the strain on the hardware to perform this on real time, so the volume of this universe might be limited.
    \item {Gravitational Lensing}. An ambitious proposal, where the weak gravitational lensing of large clusters could be visually shown. Obviously solving the full equations from General Relativity would not be viable, but it might be possible to develop a ‘'lens'’ object that acts as a close approximation.
    \item {Gamification}. As previously mentioned, Astera is a potentially invaluable outreach tool for increasing public awareness of the large scale universe. Gamifing Astera by introducing elements that make exploring the cosmological volume fun and educational could increase this value even further.
    
\end{itemize}

More information, visual materials and videos can be found on the Astera website \url{https://astera.soton.ac.uk}. {We will also update this website with any future details of Astera's public release.}


\vspace{6pt} 



\authorcontributions{Conceptualization, methodology, software, visualization, writing---original draft preparation; C. Marsden; methodology, supervision, project administration, writing---review and editing, funding acquisition; F. Shankar. All authors have read and agreed to the published version of the manuscript.} 

\funding{C. Marsden acknowledges the ESPRC funding for his PhD.  F. Shankar acknowledges partial support from a Leverhulme Trust Research Fellowship. This project has benefited from an STFC IAA Grant. We also acknowledge the benefit of an NVidia GPU grant.}

\acknowledgments{Special thanks to Ciera Sargent, Oliwia Krupa and the Nuffield Foundation. Also special thanks to Mariangela Bernardi and Peter Berhoozi for discussions on this project. We thank Miguel Aragon for helpful discussions and his excellent video that provided the inspiration for this work. We thank the Bolshoi and MultiDark simulations for their data, and the Sloan Digital Sky Survey for their galaxy imagery. This project has made extensive use of Unreal Engine 4.13 and associated development tools. We acknowledge extensive use of the Python libraries astropy, matplotlib, numpy, pandas, and scipy.}

\conflictsofinterest{ The authors declare no conflict of interest. The funders had no role in the design of the study; in the collection, analyses, or interpretation of data; in the writing of the manuscript, or in the decision to publish the results.} 

\abbreviations{The following abbreviations are used in this manuscript:\\

\noindent 
\begin{tabular}{@{}ll}
$\Lambda$CDM & $\Lambda$ Cold Dark Matter\\
HMSM & Halo Mass-Stellar Mass (relation)\\
AGN & Active Galactic Nucleus/Nuclei\\
SDSS & Sloan Digital Sky Survey\\
HST & Hubble Space Telescope\\
NFW & Navarro–Frenk–White (profile) \\
GUI & Grapical User Interface\\
RBG & Red, Green, Blue\\
GIMP & GNU Image Manipulator Program\\
FPS & Frames Per Second\\
CPU & Central Processing Unit \\
GPU & Graphics Processing Unit \\
UE4 & Unreal Engine 4\\
\end{tabular}}



\reftitle{References}

\externalbibliography{yes}
\bibliography{Bibliography}




\end{document}